\documentclass[conference]{IEEEtran}

\usepackage{amsmath,amssymb,amsfonts}
\usepackage{algorithm}
\usepackage{algorithmic}
\usepackage{cite}
\usepackage{url}
\usepackage{amsthm}
\usepackage{soul}
\usepackage{xcolor}
\usepackage{graphicx}
\usepackage{subcaption} 

\usepackage[top=0.7in, bottom=1.07in, left=0.7in, right=0.7in]{geometry}

\title{Altitude-Dependent RSRP and RSRQ Trade-offs in 5G NR UAV Networks}

\IEEEoverridecommandlockouts
\author{
\IEEEauthorblockN{Md Sharif Hossen, Vijay K. Shah and Ismail Guvenc}
\IEEEauthorblockA{Department of Electrical and Computer Engineering\\
North Carolina State University, Raleigh, NC, USA\\
Email: \{mhossen, vijay.shah, iguvenc\}@ncsu.edu
}
\thanks{This work is supported in part by the NASA ULI award 80NSSC25M7102 and in part by NSF CNS-2443035.}
}
\begin{document}
\maketitle
\begin{abstract}
Cellular-connected unmanned aerial vehicles (UAVs) in 5G NR networks experience propagation and interference conditions that vary significantly with altitude and differ substantially from those experienced by terrestrial users. This is primarily caused by the down-tilted antenna sectors in 5G NR networks, which cause UAVs to be served (and interfered with) by the sidelobes. In this paper, we develop a 3GPP-compliant system-level framework for the consistent characterization of key performance indicators (KPIs) such as reference signal received power (RSRP), reference signal received quality (RSRQ), and signal-to-interference-and-noise ratio (SINR) in a multi-site tri-sector deployment with realistic antenna patterns and probabilistic models for line-of-sight (LOS) and non-LOS (NLOS) conditions. Simulation results demonstrate that a critical transition for aerial users is experienced when going from coverage-limited to interference-limited conditions at higher altitudes. Although RSRP is affected by large-scale propagation characteristics and degrades gradually with increasing altitude and inter-site distance (ISD), SINR degrades much faster due to increased interference caused by LOS conditions. On the contrary, increasing ISD improves SINR and RSRQ due to lower interference, even as received power is reduced. 
\end{abstract}
\section{Introduction and Related Works}
The integration of unmanned aerial vehicles (UAVs) into cellular networks has gained significant attention due to their potential in applications such as surveillance, emergency response \cite{Mozaffari8660516}, and data collection \cite{hossen2026resilientuavdatamule,sharif11488226}. In particular, 5G New Radio (NR) networks provide a promising platform for enabling wide-area UAV command and control (C2) connectivity using existing terrestrial infrastructure. However, unlike ground user equipments (GUE) that operate in a two-dimensional plane, UAVs operate in a three-dimensional (3D) space, where propagation characteristics and antenna patterns vary significantly with altitude. This fundamentally alters both coverage and interference behavior in cellular networks.

In parallel, electronic conspicuity (EC) frameworks have emphasized the need for reliable detection and communication of aerial vehicles for safe airspace integration \cite{nassif2024ec}. However, existing EC approaches often suffer from coverage and interference limitations, particularly in dense environments. These challenges motivate the use of cellular infrastructure, such as 5G NR, for reliable UAV connectivity, where performance critically depends on altitude-dependent radio conditions.


As UAV altitude increases, the probability of line-of-sight (LOS) with terrestrial base stations (BSs) increases, leading to increased LOS visibility, which alters received signal characteristics. Also, it exposes UAVs to a larger number of interfering cells, significantly elevating inter-cell interference (ICI) \cite{Maeng11003933}. UAVs are often served by BS sidelobes, leading to weak desired signal strength and strong interference from multiple cells \cite{wang2022bscoop}. Hence, UAV communication transitions from coverage-limited operation at low altitudes to interference-limited operation at higher altitudes. This altitude-dependent coupling between coverage and interference presents a key challenge for reliable aerial connectivity.

Recent studies have investigated UAV communications using analytical, measurement-based, and data-driven approaches. Foundational works have characterized UAV integration into cellular networks and highlighted key challenges in coverage and interference \cite{mozaffari2019tutorial,fotouhi2019survey}, including practical UAV-assisted BS deployments that demonstrate coverage benefits and also reveal significant air--ground interference limitations \cite{haps_9448899,matraccia2021coverage}. Standardization efforts and system-level evaluations based on 3GPP models provide realistic frameworks for aerial users \cite{3gpp38901,3GPP36777,matraccia2021coverage}. Measurement-driven studies have revealed altitude-dependent propagation and interference characteristics \cite{Maeng11003933,Jie10587227}, including empirical channel characterization works that show reduced multipath effects and path loss variations with increasing altitude \cite{liu2019measurement}. In addition, comprehensive surveys on UAV-enabled networks highlight key challenges in communication reliability and network integration, but primarily provide qualitative insights rather than detailed system-level modeling \cite{uav_iot_survey}. Recent works have also analyzed the impact of LOS-dominated interference at higher altitudes \cite{gapeyenko2018analysis,zeng2019accessing}. Data-driven approaches have also been explored for UAV channel modeling and performance prediction \cite{Alobaidy11269020}.

However, most existing works analyze performance metrics such as reference signal received power (RSRP) or signal-to-interference-and-noise ratio (SINR) independently, or without a unified framework that consistently captures the interplay between received power, interference, and noise across KPIs such as RSRP, reference signal received quality (RSRQ), and SINR. Moreover, while antenna patterns are often implicitly considered, the role of downtilted BS antennas and the resulting sidelobe-dominated connectivity for aerial users is not explicitly analyzed at the system level, which brings a gap in the altitude-dependent interference behavior of cellular networks. This work addresses these limitations by explicitly modeling sidelobe-driven connectivity and multi-cell interference for aerial users in 3GPP-compliant multi-cell 5G NR networks. By explicitly coupling received power, aggregate interference, and noise within a unified formulation, the proposed framework enables a consistent and physically interpretable analysis of altitude-dependent coverage–interference trade-offs, which is critical for the design and optimization of aerial cellular networks, where interference-aware mechanisms are required to ensure reliable UAV connectivity.

Motivated by the recent elevated interest in serving UAVs by 5G NR networks \cite{nassif2024ec}, the main contributions of this paper are summarized as follows:
\begin{itemize}
\item We develop a 3GPP-compliant system-level framework for evaluating altitude-dependent UAV performance in 5G NR networks, incorporating realistic deployment geometry, antenna patterns, interference conditions, and probabilistic LOS/NLOS propagation.
    \item We establish a consistent KPI formulation that jointly captures received power, interference, and noise, ensuring alignment across RSRP, RSRQ, and SINR.
    
    \item We characterize the altitude-dependent transition from coverage-limited to interference-limited operation and quantify its impact on aerial link performance.
    
    \item We reveal a non-intuitive inter-site distance (ISD) trade-off, where increasing ISD reduces interference and improves SINR and RSRQ despite degrading RSRP.
    \item We demonstrate that RSRQ serves as a practical indicator of interference dominance and captures the coverage–interference decoupling in aerial networks.
\end{itemize}

The remainder of this paper is organized as follows. Section~\ref{sec:system_model} presents the system model, including the network topology, antenna configuration, and propagation modeling. Section~\ref{sec:metrics_kpi-charact} defines the performance metrics and KPI characterization. Section~\ref{sec:results} provides numerical results and discusses key observations on altitude-dependent coverage and interference trade-offs. Finally, Section~\ref{sec:conclusion} concludes the paper.






\section{System Model}
\label{sec:system_model}
In this section, we discuss the cellular deployment geometry,  antenna pattern, and channel models.

\subsection{Network Topology}
\begin{figure}[t]
\centering  \includegraphics[width=0.9\linewidth]{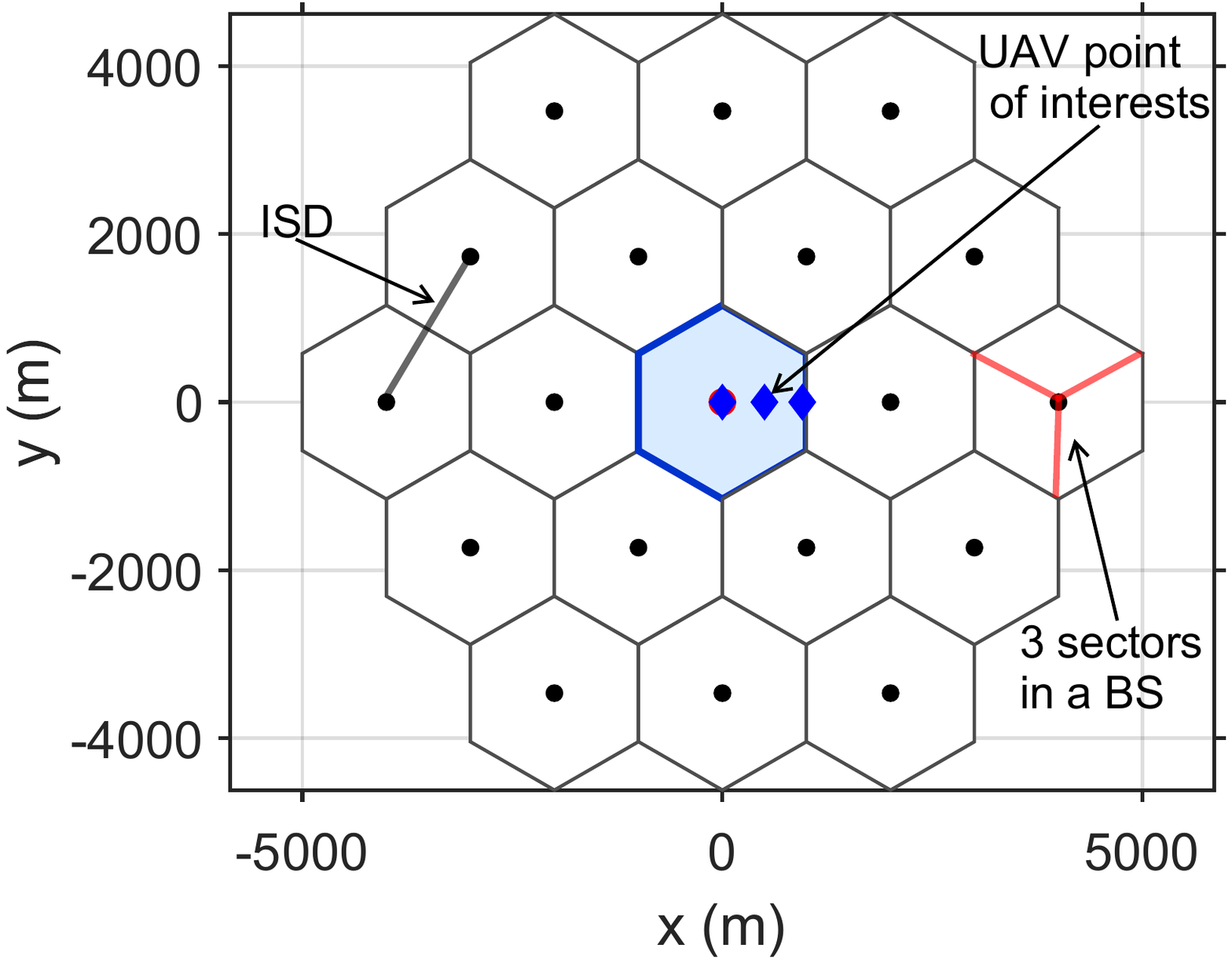}
\caption{A 19-cell hexagonal cellular
deployment with tri-sector BSs with wrap-around geometry.
}
\vspace{-0.12in}
\label{fig:sys_geom}
\end{figure}

We consider a 19-cell hexagonal cellular deployment with inter-site distance ($\mathrm{ISD}$) as a control parameter to study performance trade-offs. Each base station (BS) site is equipped with three sectors separated by $120^\circ$ in azimuth. This results in a total of $N=57$ sectors, and each sector is indexed by $s \in \{1,2,\dots,N\}$.
Fig.~\ref{fig:sys_geom} shows the network geometry where the center cell is used as the region for UAV performance evaluation, while the surrounding cells generate ICI. To ensure boundary effects and consistent interference, wrap-around geometry is employed so that edge users experience interference equivalent to that in an infinite cellular layout \cite{3gpp38901} \cite{3GPP36777}.

\subsection{Aerial User Geometry}
The UAV is placed at an altitude $h$ to operate within the center cell of the network, where three deterministic horizontal positions $(x,y)$ are considered, which are located at cell-center, cell-middle, and cell-edge, respectively. This allows aerial user locations to experience different antenna gains and interference conditions from multiple sectors. Let the UAV position is $\mathbf{u} = [x, y, h]^T$ and the two- and three-dimensional distance between sector $s$ and the UAV for a BS height $h_b$ is:
\begin{equation}
\begin{aligned}
d_\mathrm{2D}&= \sqrt{(x-x_s)^2 + (y-y_s)^2 }\\
\quad d_\mathrm{3D} &= \sqrt{d_\mathrm{2D}^2 + (h-h_{\mathrm{b}})^2}
\end{aligned}
\label{eq:distance}
\end{equation}
where the sector's horizontal position is denoted by $(x_s, y_s)$. Then, elevation $(\theta_s)$ and azimuth $(\phi_s)$ angles with respect to $b$ are given by:
\begin{equation}
\begin{aligned}
\theta_s = \tan^{-1}\!\left(
\frac{h-h_{\mathrm{b}}}{d_\mathrm{3D}}
\right),~~
\phi_s = \tan^{-1}\!\left(\frac{y - y_s}{x - x_s}\right) 
\end{aligned}
\end{equation}

\begin{figure*}[t]
\centering
\includegraphics[width=.9\textwidth,trim=0in 1in 0in .75in,clip]
{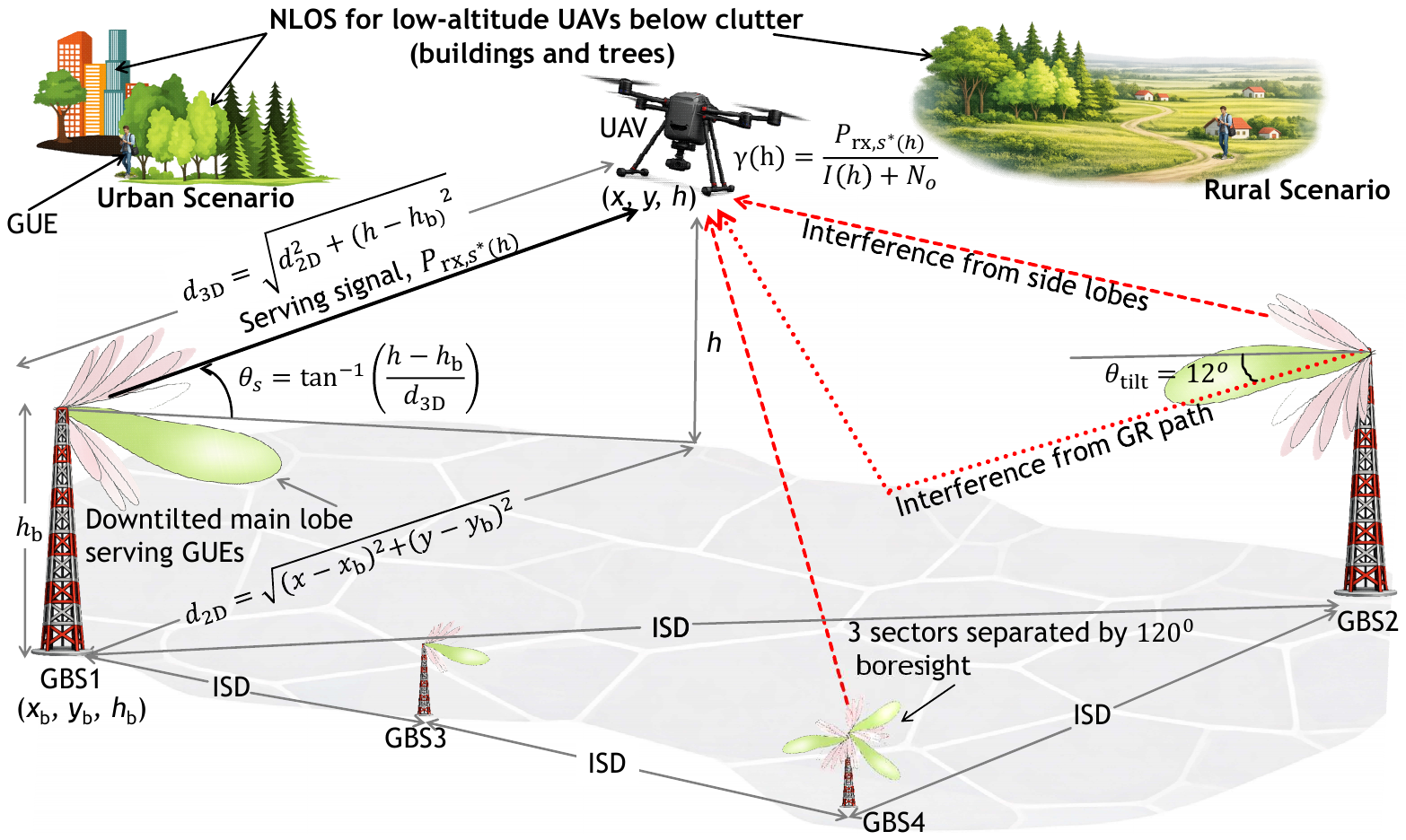}
\caption{Geometry of UAV-BS links illustrating 3D distance, elevation angle, and interference from neighboring sectors in urban and rural environments.}
\label{fig:geometry_uav}
\vspace{-.12in}
\end{figure*}
Fig.~\ref{fig:geometry_uav} illustrates the UAV–BS link. Unlike Fig.~\ref{fig:sys_geom}, which shows the overall network layout, Fig.~\ref{fig:geometry_uav} focuses on the link-level relationships between the UAV and surrounding BS sectors. 
Along with the elevation angle, it also illustrates the down-tilted antenna pattern, where the main lobe primarily serves GUEs, while UAVs, as aerial users, are often served through sidelobes. As a result, the UAV is exposed to strong interference from multiple neighboring sectors, including both direct sidelobe radiation and ground-reflected (GR) components. Additionally, low-altitude UAVs often operate in NLOS conditions due to buildings in urban environments and trees in rural environments.
The figure also provides intuition for the trade-off between received signal strength and signal quality for aerial users. While higher UAV altitudes can improve RSRP due to reduced blockage and an increased likelihood of LOS links, they also expose the UAV to more interfering sectors, degrading RSRQ. This trade-off is a key characteristic of aerial cellular communication and motivates the need for careful analysis of altitude-dependent performance.

\subsection{Sector Antenna Pattern}
Each BS sector employs the 3GPP composite antenna model \cite{3gpp38901},
which characterizes directional gain in both the horizontal and
vertical planes. The antenna is defined by the electrical downtilt
$\theta_{\mathrm{tilt}}$, the vertical and horizontal $3$~dB beamwidths
$\theta_{3\mathrm{dB}}$ and $\phi_{3\mathrm{dB}}$, and the maximum gain
$G_{\max}$.

Let $\psi_s$ denote the azimuth boresight direction of sector $s$.
The azimuth and elevation offsets between the UAV direction and the
sector boresight are given by:
\begin{equation}
\Delta\phi_s(h)=\phi_s(h)-\psi_s, \qquad
\Delta\theta_s(h)=\theta_s(h)+\theta_{\mathrm{tilt}} .
\end{equation}
The vertical and horizontal attenuations are given by:
\begin{equation}
A_{v,s}(h)=\min\!\left(\mathrm{SLA}_v,
12\left(\frac{\Delta\theta_s(h)}{\theta_{3\mathrm{dB}}}\right)^2\right),
\end{equation}
\begin{equation}
A_{h,s}(h)=\min\!\left(A_m,
12\left(\frac{\Delta\phi_s(h)}{\phi_{3\mathrm{dB}}}\right)^2\right),
\end{equation}
where $\mathrm{SLA}_v$ is the vertical sidelobe attenuation limit and
$A_\mathrm{m}$ is the maximum attenuation. The total attenuation for sector $s$, $A_s(h)$, toward the UAV and antenna gain, $G_s(h)$, at $h$ is:
\begin{equation}
A_s(h)=\min\!\left(A_m,\,A_{v,s}(h)+A_{h,s}(h)\right),
\end{equation}
\begin{equation}
G_s(h)=G_{\max}-A_s(h)\;(\mathrm{dBi}),
\end{equation}

\subsection{Propagation and Channel Modeling}
Large-scale propagation between the UAV and cellular sectors is modeled with 3GPP TR~38.901 \cite{3gpp38901}. The received signal attenuation consists of distance-dependent path loss, probabilistic LOS or non-NLOS propagation, and log-normal shadow fading. We consider urban macrocell (UMa) and rural macrocell (RMa) environments.

\subsubsection{UMa Path Loss}
For urban deployments, LOS propagation is modeled using a two-slope path-loss model with a breakpoint at a specified distance. This structure captures the transition from a free-space dominated propagation region to a region influenced by ground-reflected components: 
\begin{equation}
\label{eq:L_UMA_FS}
L_{\mathrm{FS}}^{\mathrm{UMa}}
=28 + 22\log_{10}(d_\mathrm{3D}) + 20\log_{10}(f_\mathrm{c}),    
\end{equation}
where $f_\mathrm c$ is the carrier frequency, and \(L_{\mathrm{FS}}^{\mathrm{UMa}}\) represents the free-space-dominated attenuation for links whose horizontal distance $d_\mathrm{2D}$ is below the breakpoint distance, $d_{\mathrm{BP}}^{\mathrm{UMa}}$. Beyond $d_{\mathrm{BP}}^{\mathrm{UMa}}$, the path loss is:
\begin{equation}
\begin{aligned}
L_\mathrm{GR}^{\mathrm{UMa}}
&=28 + 40\log_{10}(d_\mathrm{3D}) + 20\log_{10}(f_c)\\
& \quad - 9\log_{10}\!\left( \left(d_{\mathrm{BP}}^{\mathrm{UMa}}\right)^2 + \left(h_\mathrm{b}-h\right)^2 \right), 
\end{aligned}
\end{equation}
where \(L_\mathrm{GR}^{\mathrm{UMa}}\) captures the stronger attenuation caused by the two-ray ground reflection effect that becomes dominant at longer link distances. The $d_\mathrm{BP}^{\mathrm{UMa}}$ is defined as:
\begin{equation}
d_\mathrm{BP}^{\mathrm{UMa}}
=
\frac{4 (h_\mathrm{b}-h_e)(h-h_e) f_\mathrm{c}}{c},    
\end{equation}
where $c$ is the speed of light, and $h_\mathrm{e} = 1$ m represents the effective environment average height of surrounding scatterers. Then, LOS path loss for UMa is given by:
\begin{equation}
\label{eq:los_UMa}
L_{\mathrm{LOS}}^{\mathrm{UMa}} =
\begin{cases}
L_\mathrm{FS}^{\mathrm{UMa}}, & d_\mathrm{2D} \le d_\mathrm{BP}^{\mathrm{UMa}} \\
L_\mathrm{GR}^{\mathrm{UMa}}, & d_\mathrm{2D} > d_\mathrm{BP}^{\mathrm{UMa}}.
\end{cases}
\end{equation}
Hence, \eqref{eq:los_UMa} captures the propagation behavior for both short-range and long-range LOS links. For UMa NLOS propagation, the path loss is given by:
\begin{equation}
L_{\mathrm{NLoS}}^{\mathrm{UMa}}
=\max\!\left(
L_{\mathrm{LoS}}^{\mathrm{UMa}},
L_{\mathrm{NLoS}}^{\mathrm{UMa}^\prime}
\right),    
\end{equation}
where
\begin{equation}
\begin{aligned}
L_{\mathrm{NLOS}}^{\mathrm{UMa}^\prime}
&=
13.54
+ 39.08\log_{10}(d_\mathrm{3D})
+ 20\log_{10}(f_\mathrm c)\\
&\quad- 0.6(h-1.5).
\end{aligned}
\end{equation}

\subsubsection{RMa Path Loss}
For RMa, the breakpoint distance is:
\begin{equation}
d_\mathrm{BP}^{\mathrm{RMa}} = \frac{2\pi~h_\mathrm{b} h f_\mathrm{c}}{c},    
\end{equation}
Hence, the LOS path loss for RMa is given by:
\begin{equation}
L_{\mathrm{LOS}}^{\mathrm{RMa}} =
\begin{cases}
L_\mathrm{FS}^{\mathrm{RMa}}, & d_\mathrm{2D} \le d_\mathrm{BP}^{\mathrm{RMa}} \\
L_\mathrm{GR}^{\mathrm{RMa}}, & d_\mathrm{2D} > d_\mathrm{BP}^{\mathrm{RMa}}
\end{cases}    
\end{equation}
where for the average building height $h_\mathrm{blg} = 5$ m,
\begin{equation}
\begin{aligned}
L_\mathrm{FS}^{\mathrm{RMa}} &=
20\log_{10}\!\left(\frac{40\pi~  d_\mathrm{3D} f_\mathrm c}{3}\right)
+ \min(0.03h^{1.72},10) \\ & \quad \times \log_{10}(d_\mathrm{3D})
- \min(0.044h_\mathrm{blg}^{1.72},14.77)\\
&\quad+ 0.002\log_{10}(h)d_\mathrm{3D} 
\end{aligned}
\end{equation}

and
\begin{equation}
L_\mathrm{GR}^{\mathrm{RMa}} =
L_\mathrm{FS}^{\mathrm{RMa}}
+ 40\log_{10}\left(\frac{d_\mathrm{3D}}{d_\mathrm{BP}^{\mathrm{RMa}}}\right).
\end{equation}

The RMa NLOS path loss is given by:
\begin{equation}
\label{eq:L_RMa_NLOS}
L_{\mathrm{NLOS}}^{\mathrm{RMa}}
=\max\left(L_{\mathrm{LOS}}^{\mathrm{RMa}},L_{\mathrm{NLOS}}^{\mathrm{RMa}\prime}\right),    
\end{equation}
where for average street width, $W=20$,
\begin{equation}
\begin{aligned}
L_\mathrm{NLOS}^{\mathrm{RMa}}\prime
&=161.04
- 7.1\log_{10}(W)
+ 7.5\log_{10}(h_\mathrm{bl})
\\&\quad-\left(
24.37
- 3.7\left(\frac{h_\mathrm{bl}}{h_\mathrm b}\right)^2
\right)\log_{10}(h_\mathrm{b})
\\&\quad+
\left(
43.42
- 3.1\log_{10}(h_\mathrm{b})
\right)
(\log_{10}(d_\mathrm{3D})-3)
\\&\quad+
20\log_{10}(f_\mathrm c)
-
\left(
3.2(\log_{10}(11.75h))^2
-4.97
\right)
\end{aligned}
\end{equation}

\subsubsection{LOS Probability and Propagation State}
Although the deterministic path losses are determined in~\eqref{eq:L_UMA_FS}–\eqref{eq:L_RMa_NLOS}, the propagation condition between the BS and the UAV must be determined as either LOS or NLOS. Hence, the LOS probability is modeled as a function of the horizontal distance $d_\mathrm{2D}$. For the RMa and UMa scenarios, the LOS probabilities are respectively defined as:
\begin{equation}
\label{eq:rma_prob}
P^\mathrm{RMa}_{\mathrm{LOS}}(d_\mathrm{2D}) =
\begin{cases}
1, & d_\mathrm{2D} \le 10~\text{m} \\
\exp\!\left(-\dfrac{d_\mathrm{2D}-10}{1000}\right), & d_\mathrm{2D} > 10~\text{m}~.
\end{cases}
\end{equation}
\begin{equation}
\label{eq:uma_prob}
P^\mathrm{UMa}_{\mathrm{LOS}}(d_\mathrm{2D}) =
\begin{cases}
1, & d_\mathrm{2D} \le 18~\text{m} \\
X,
& d_\mathrm{2D} > 18~\text{m}
\end{cases}
\end{equation}
where
\begin{equation}
\begin{aligned}
X&=\left(\dfrac{18}{d_\mathrm{2D}} + 
\exp\!\left(-\dfrac{d_\mathrm{2D}}{63}\right)
\left(1-\dfrac{18}{d_\mathrm{2D}}\right)\right)\\ 
& \quad \times
\left(1 + C'(h)\dfrac{5}{4}
\left(\dfrac{d_\mathrm{2D}}{100}\right)^3
\exp\!\left(-\dfrac{d_\mathrm{2D}}{150}\right)\right),    
\end{aligned}
\end{equation}
\begin{equation}
C'(h) =
\begin{cases}
0, & h \le 13~\text{m}, \\
\left(\dfrac{h-13}{10}\right)^{1.5},
& 13 < h \le 23~\text{m}.
\end{cases}
\end{equation}

However, \eqref{eq:rma_prob} and \eqref{eq:uma_prob} represent the likelihood of LOS for RMa and UMa scenarios rather than the actual propagation condition. Hence, a Bernoulli random variable is used to generate the propagation state, $Z_s$ for a sector $s$. Let $U \sim \mathcal{U}(0,1)$ be a uniformly distributed random variable. Then,
\begin{equation}
Z_s =
\begin{cases}
\text{LOS}, & U < P_{\mathrm{LOS}}(d_\mathrm{2D}) \\
\text{NLOS}, & \text{otherwise}
\end{cases}
\end{equation}
The path loss for sector $s$ is then given by:
\begin{equation}
L_s =
Z_s L_{\mathrm{LOS}}
+
(1-Z_s)L_{\mathrm{NLOS}}.   
\end{equation}

To capture large-scale signal fluctuations caused by environmental obstructions, log-normal shadow fading is applied:
\begin{equation}
X_s \sim \mathcal{N}(0,\sigma_s^2).   
\end{equation}

where $\sigma_s$ depends on the propagation state and scenario. The effective large-scale attenuation between the UAV and sector $s$ becomes:
\begin{equation}
L_s^\prime = L_s + X_s,   
\end{equation}
This probabilistic realization is repeated across Monte-Carlo trials to capture the stochastic nature of blockage conditions in aerial cellular propagation across both urban and rural environments.
\section{Performance Metrics and KPI Characterization}
In this section, we define the performance metrics and their statistical characterization for UAV communication analysis.
\label{sec:metrics_kpi-charact}
\subsection{Received Power and Best Cell Association}
The received power from each BS sector $s$ is given as:
\begin{equation}
P_{\mathrm{rx},s} =
P_{\mathrm{tx},s} + G_{s}(\theta,\phi) - L'_s - L_{\mathrm{impl}},
\end{equation}
where $P_{\mathrm{tx},s}$ is the BS transmit power, $G_{s}(\theta,\phi)$ represents the sector antenna gain toward the UAV, and $L_{\mathrm{impl}}$ accounts for implementation losses.

The RSRP $(\mathcal{R})$, for each $s$ is obtained by normalizing the received power, $P_{\mathrm{rx}}$, with respect to the number of resource elements $(N_\mathrm{RB})$:
\begin{equation}
\mathcal{R}_s = P_{\mathrm{rx},s} - 10\log_{10}(12N_\mathrm{RB}),
\end{equation}

For each UAV position, the serving sector $s^{\star}(h)$ at $h$ is selected by the maximum instantaneous $\mathcal{R}_s$:
\begin{equation}
s^{\star}(h) \triangleq \arg\max_{s\in\mathcal{S}} \mathcal{R}_s(h),
\label{eq:association}
\end{equation}

\subsection{SINR and RSRQ Computation}
Although RSRP is the main metric for cell selection and handover decisions in LTE and 5G, it does not solely capture the actual signal quality \cite{Alobaidy11269020}. Hence, signal-to-interference-plus-noise ratio (SINR) and reference signal received quality (RSRQ) measurements are necessary. SINR, $\gamma$, for each sector towards the UAV is computed in linear scale (mW) as:
\begin{equation}
\gamma_s(h) =
\frac{P_{\mathrm{rx},s{^\star}}(h)}
{I_s(h) + N_0},
\end{equation}
where $N_0$ represents the receiver noise power, and the aggregate interference at height $h$ is given by:
\begin{equation}
I_s(h) \triangleq \rho \sum_{s \ne s^\star} P_{\mathrm{rx},s}(h),
\end{equation}
where $\rho \in[0,1]$ is the interference scaling factor, RSRQ, represented by $\mathcal{Q}_s(h)$ is computed as:
\begin{equation}
\mathcal{Q}^\mathrm{dB}_s(h) = \mathcal{R}_s(h) - \mathcal{I}^\prime_s(h) + 10\log_{10}(N_\mathrm{RB}),
\end{equation}
where $\mathcal{I}^\prime_s(h)$ is the received signal strength indicator (RSSI), which accounts for the total received signal strength, including the serving signal, interference, and noise, and is given by:
\begin{equation}
\mathcal{I}^\prime_s(h) \triangleq P_{\mathrm{rx},s^\star}(h) + I_s(h) + N_0
\end{equation}

\subsection{Statistical KPI Characterization}
The objective of this work is to characterize the altitude-dependent behavior of key cellular performance metrics for UAVs in a multi-cell 5G NR network with particular emphasis on the effect of terrestrial BS downtilt and sidelobe-driven aerial coverage. Specifically, for a given UAV position $\mathbf{u}$
, we evaluate the statistical behavior of RSRP, RSRQ, and SINR as functions of altitude for the serving sector given by:
\begin{equation}
R(h) \triangleq R_{s^\star}(h), \quad 
Q(h) \triangleq Q_{s^\star}(h), \quad 
\gamma(h) \triangleq \gamma_{s^\star}(h),
\end{equation}
under realistic propagation, antenna, and interference conditions. Since UAVs are often served through sidelobes of downtilted BS antennas while simultaneously experiencing interference from multiple neighboring sectors, these metrics exhibit a strong altitude dependence. Due to the stochastic nature of LOS/NLOS propagation, shadow fading, antenna, and sidelobe effects, the resulting KPIs are random variables. Therefore, we focus on their statistical characterization across Monte Carlo realizations, including mean and percentile (not shown due to page limit) behavior as functions of altitude. 

\section{Numerical Results}
In this section, we present results and evaluate altitude-dependent trade-offs across RSRP, SINR, and RSRQ.
\label{sec:results}
\subsection{Simulation Setup}
\label{subsec:sim_setup}

Table~\ref{tab:sim_params} summarizes the key simulation parameters used throughout the evaluation. A multi-site, tri-sector cellular deployment is considered, with all sites operating on the same carrier frequency and employing identical antenna configurations. For each altitude and horizontal distance, multiple Monte Carlo trials are performed to capture the effects of shadow fading and cell selection variability.

\begin{table}[t]
\caption{Simulation Parameters}
\label{tab:sim_params}
\centering
\begin{tabular}{ll}
\hline
Parameter & Value / Notation \\
\hline
$f_c$ & 3.5 GHz \\
NR operating band & FR1 \\
$d^\mathrm{UMa}_\mathrm{BP}$ & Breakpoint distance for UMa LOS model\\
$d^\mathrm{RMa}_\mathrm{BP}$ & Breakpoint distance for RMa LOS model\\
Channel bandwidth / SCS & 20 MHz / 30 kHz \\
$N_{\mathrm{RB}}$ & 51 \\
SSB RBs & 20 \\
$P_\mathrm{tx}$ & 46 dBm \\
Noise figure & 7 dB \\
$L_{\mathrm{impl}}$ & 8 dB \\
$G_\mathrm{max}$ & 17 dBi \\
$\phi_{3\mathrm{dB}}$~,~$\theta_{3\mathrm{dB}}$ & $65^\circ$,$65^\circ$ \\
$\theta_\mathrm{tilt}$ & $12^\circ$ \\
SLA\_$v$ & 30 dB \\
$A_\mathrm m$ & 30 dB \\
$\psi_s$ & $[0^\circ,120^\circ,240^\circ]$ \\
$h_\mathrm b$ & 30 m \\
ISD & 500, 1000, 1500, 2000 m \\
$h$ & 10, 25, 50, 100, 150, 300 m \\
Monte Carlo trials & 200 \\
$\sigma$ (UMa) & $\sigma_{\mathrm{LOS}} = 4$ dB, $\sigma_{\mathrm{NLOS}} = 6$ dB \\
$\sigma$ (RMa) & $\sigma_{\mathrm{LOS}} = 4$ dB, $\sigma_{\mathrm{NLOS}} = 8$ dB \\ 
$\rho$ & 1 \\
\hline
\end{tabular}
\vspace{-0.15in}
\end{table}

\subsection{Coverage and Signal Strength Analysis}
\begin{figure}[t]
\centering
\subfloat[RMa scenario]{%
\includegraphics[width=0.48\linewidth]{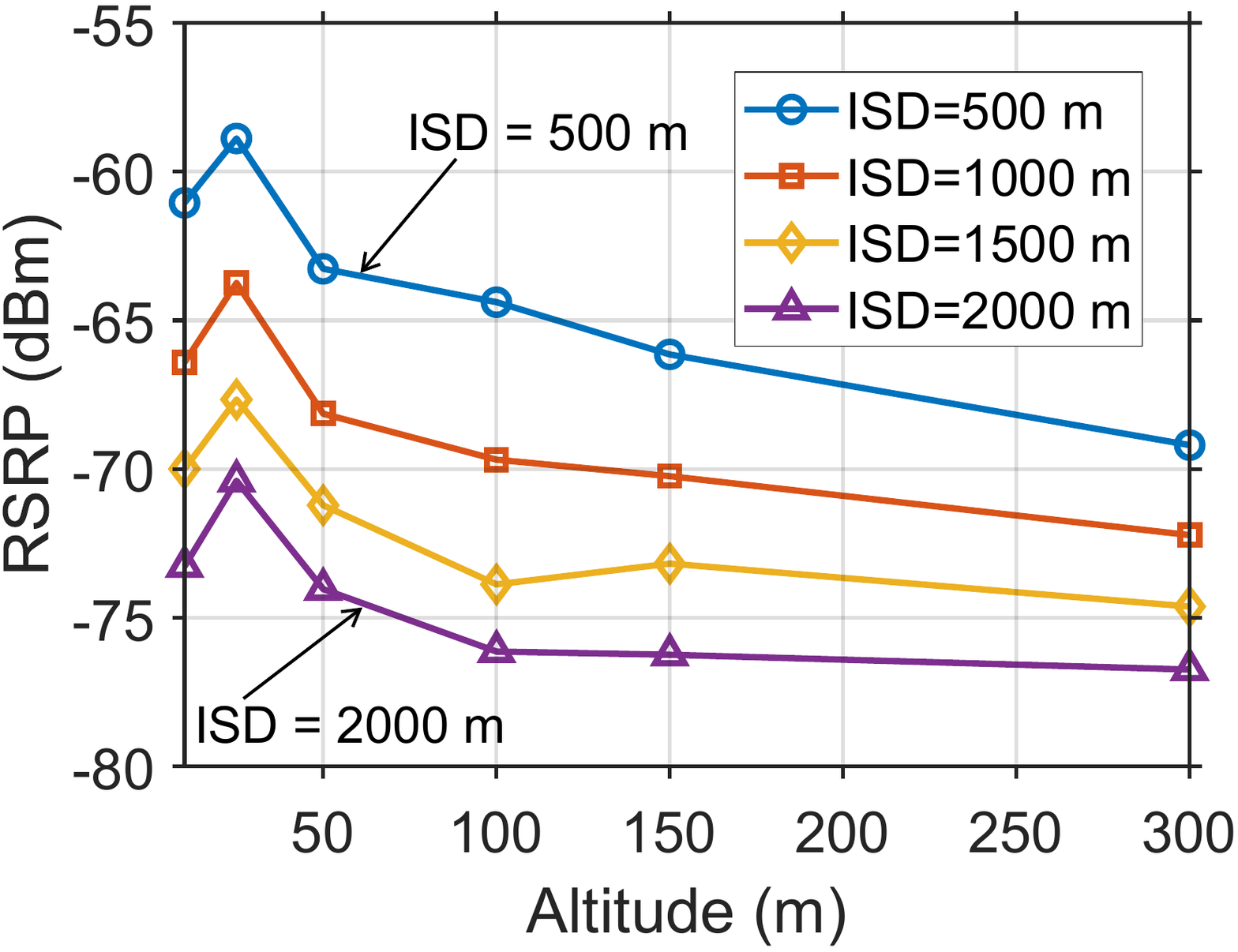}
}
\hfill
\subfloat[UMa scenario]{%
\includegraphics[width=0.48\linewidth]{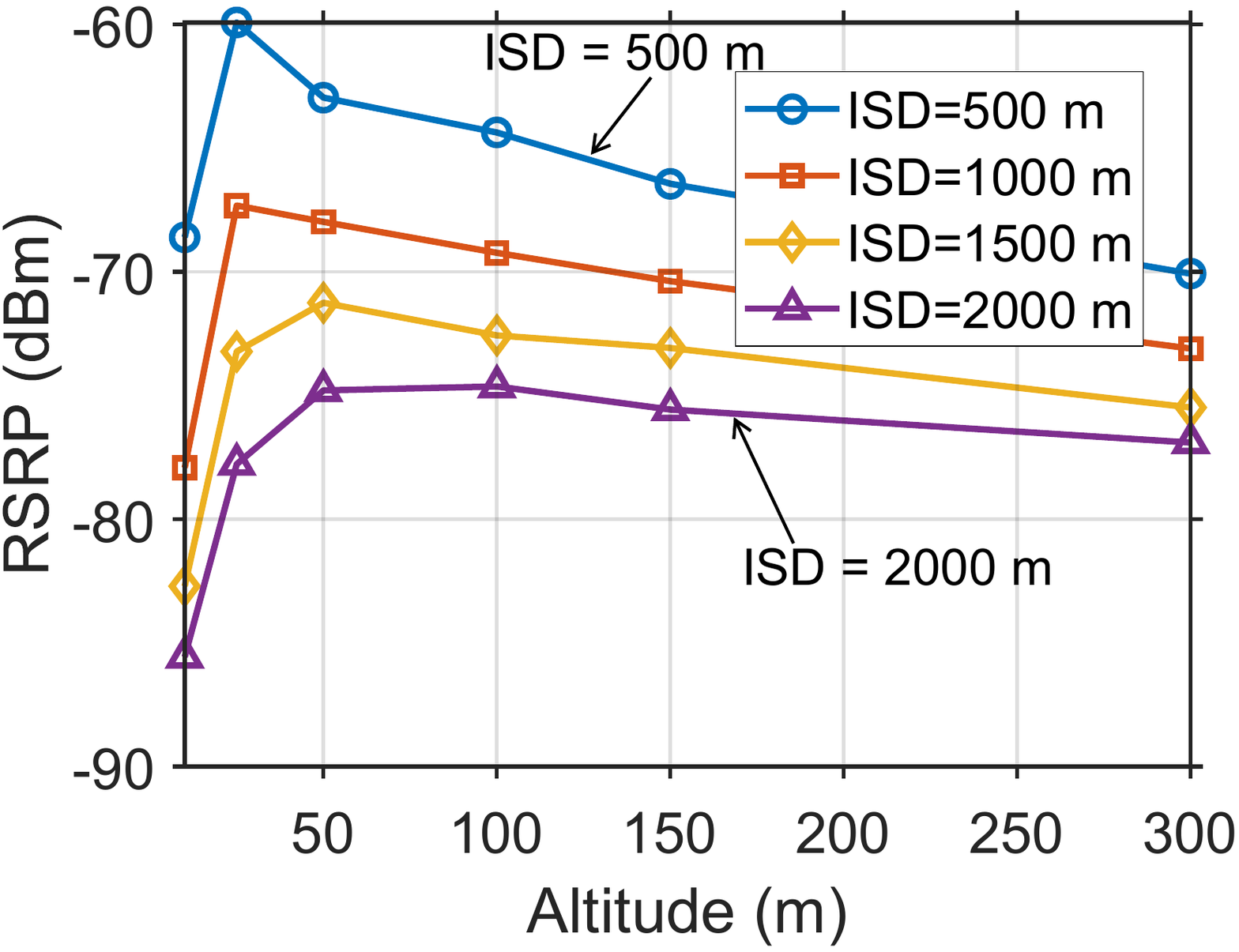}
}
\caption{Mean RSRP versus UAV altitude for different ISDs.}
\label{fig:rsrp_alt}
\vspace{-.2in}
\end{figure}

As shown in Fig.~\ref{fig:rsrp_alt}, RSRP decreases gradually with altitude due to increasing path loss and reduced effective antenna gain, as UAVs are primarily served through sidelobes of downtilted BS antennas. At low altitudes, relatively stronger RSRP is observed due to shorter link distances and partial antenna gain despite sidelobe-based coverage.


\subsection{Interference and SINR Analysis}

\begin{figure}[!h]
\centering
\subfloat[RMa scenario]{%
\includegraphics[width=0.48\linewidth]{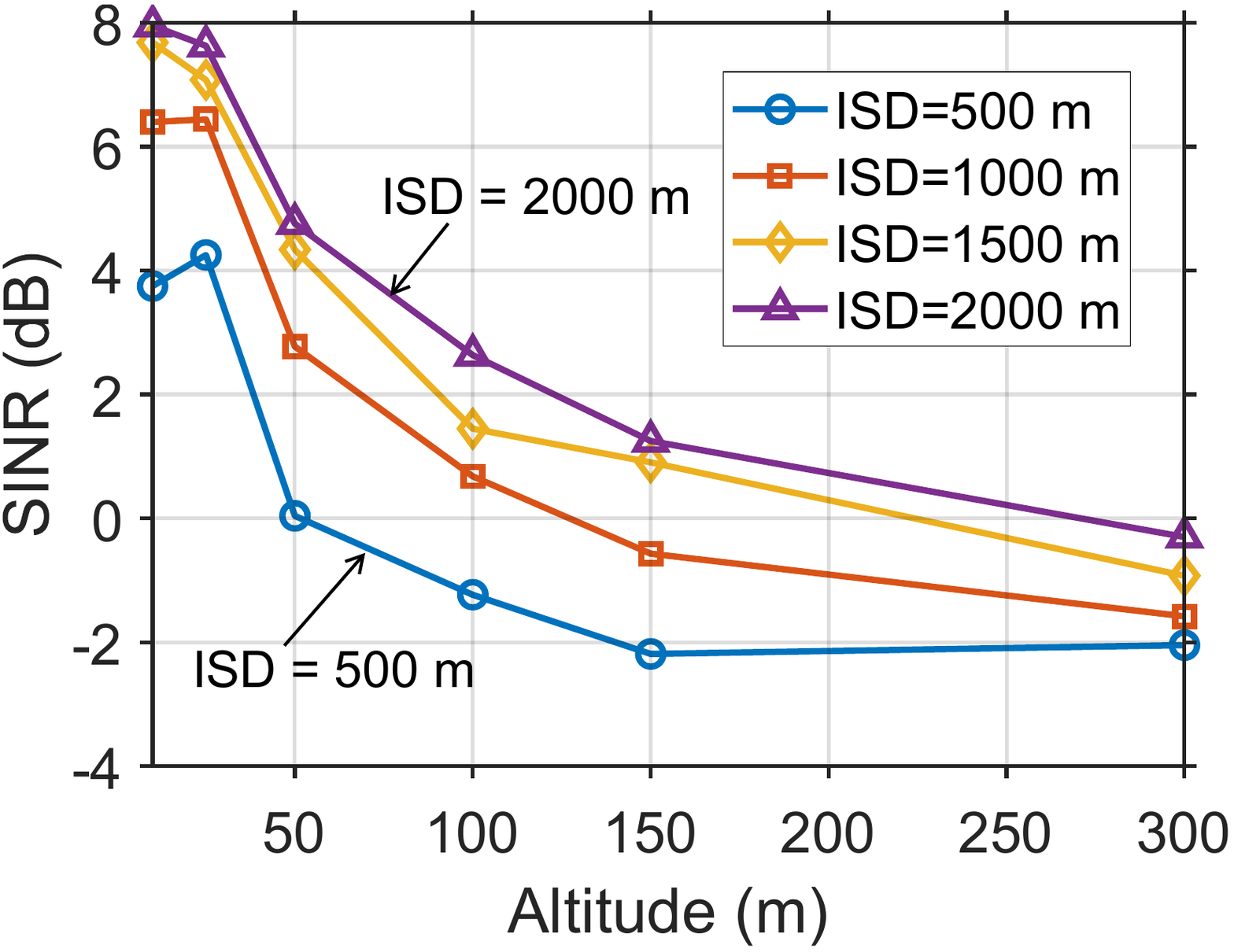}
}
\hfill
\subfloat[UMa scenario]{%
\includegraphics[width=0.48\linewidth]{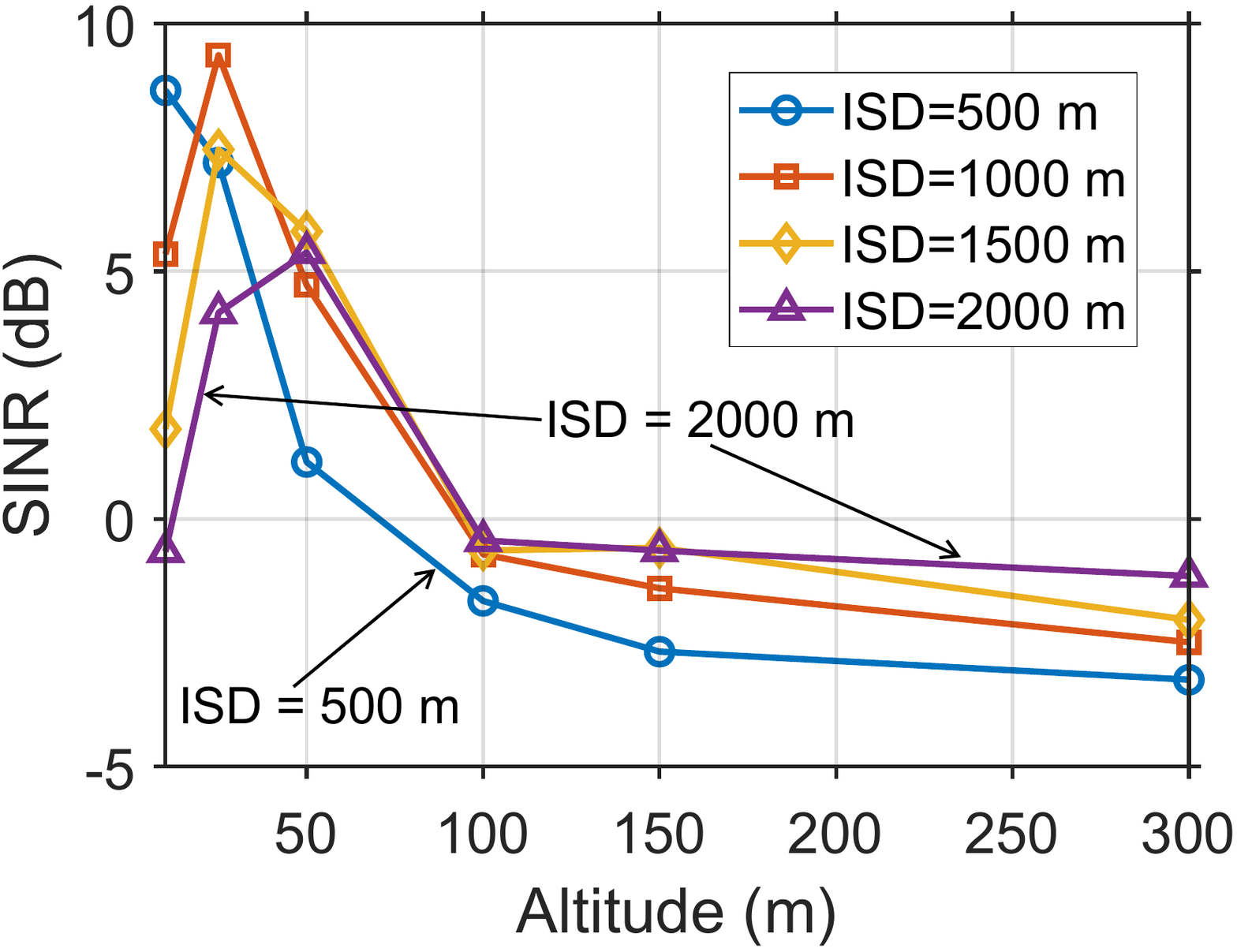}
}
\caption{Mean SINR versus UAV altitude for different ISDs.}
\label{fig:sinr_alt}
\vspace{-.12in}
\end{figure}

Fig.~\ref{fig:sinr_alt} shows SINR as a function of altitude for different ISD values. Unlike RSRP, SINR degrades rapidly with altitude due to increased LOS visibility to multiple BSs, which significantly elevates ICI.

\subsection{RSRQ Analysis}

\begin{figure}[!h]
\centering
\subfloat[RMa scenario]{%
\includegraphics[width=0.48\linewidth]{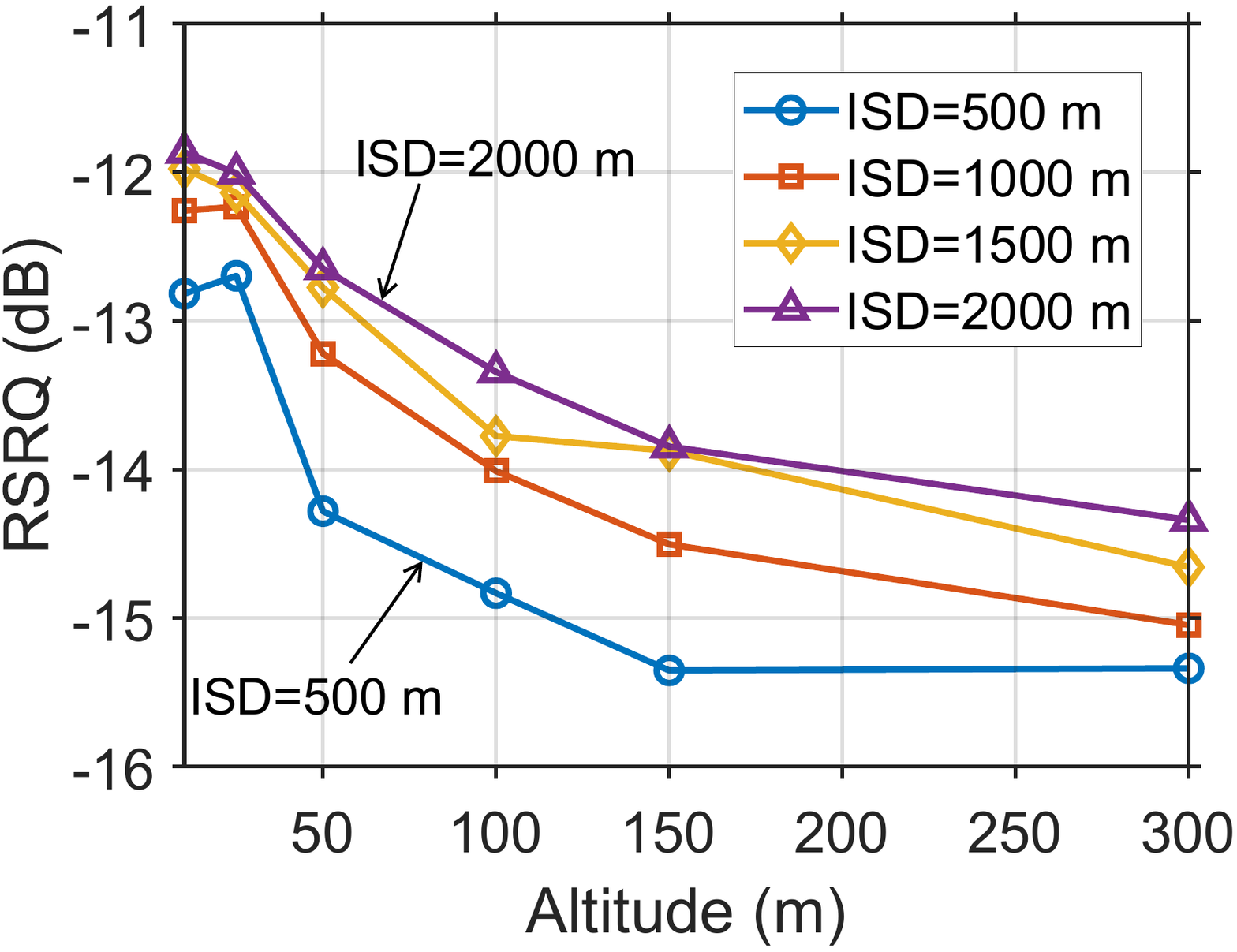}
}
\hfill
\subfloat[UMa scenario]{%
\includegraphics[width=0.48\linewidth]{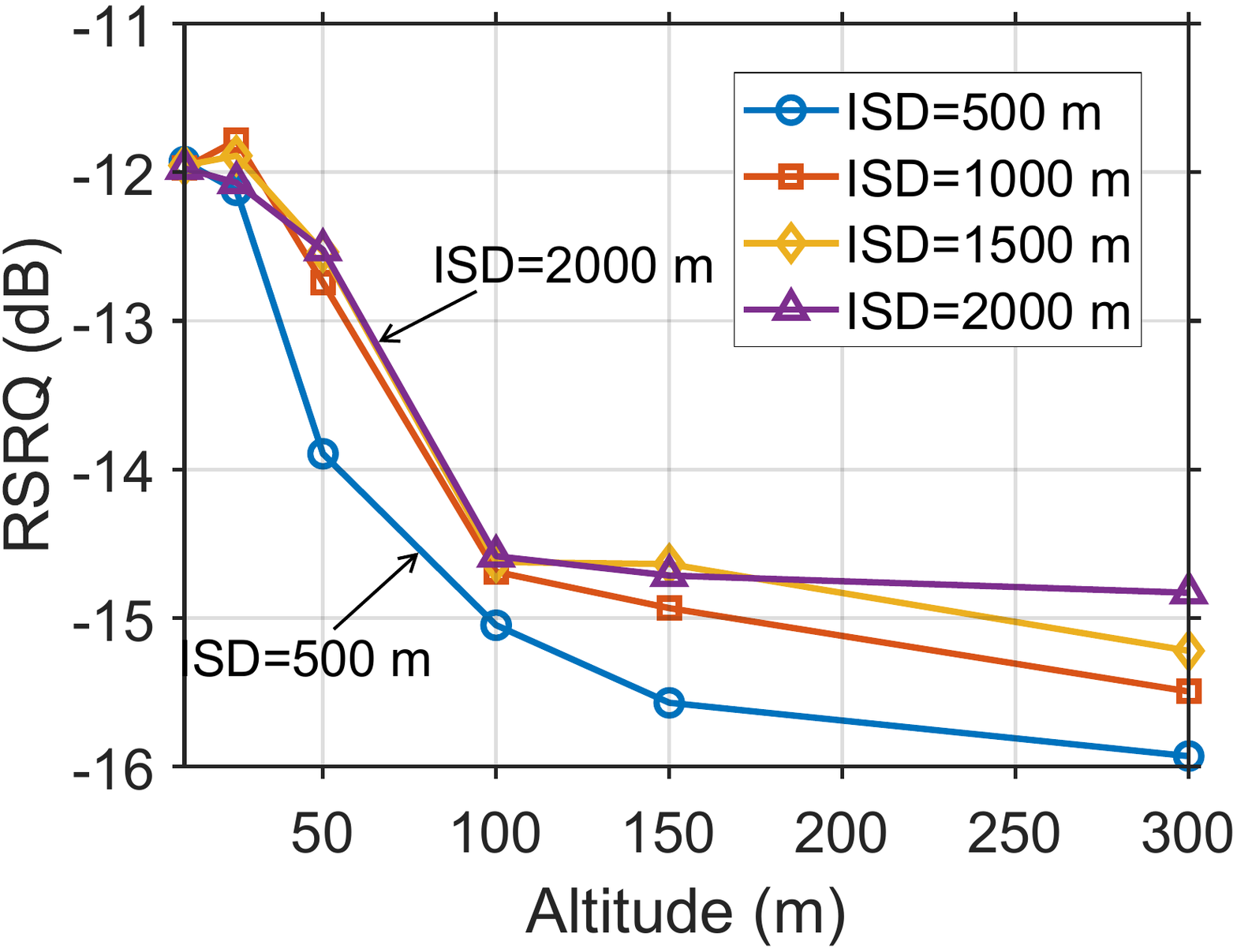}
}
\caption{Mean RSRQ versus UAV altitude for different ISDs.}
\label{fig:rsrq_alt}
\vspace{-.2in}
\end{figure}

As shown in Fig.~\ref{fig:rsrq_alt}, RSRQ decreases with altitude as interference grows faster than the desired signal power, reflecting the transition toward interference-limited operation.

\subsection{Coverage--Quality Tradeoff Analysis}
\begin{figure}[!h]
\centering
\subfloat[RMa scenario]{%
\includegraphics[width=0.48\linewidth]{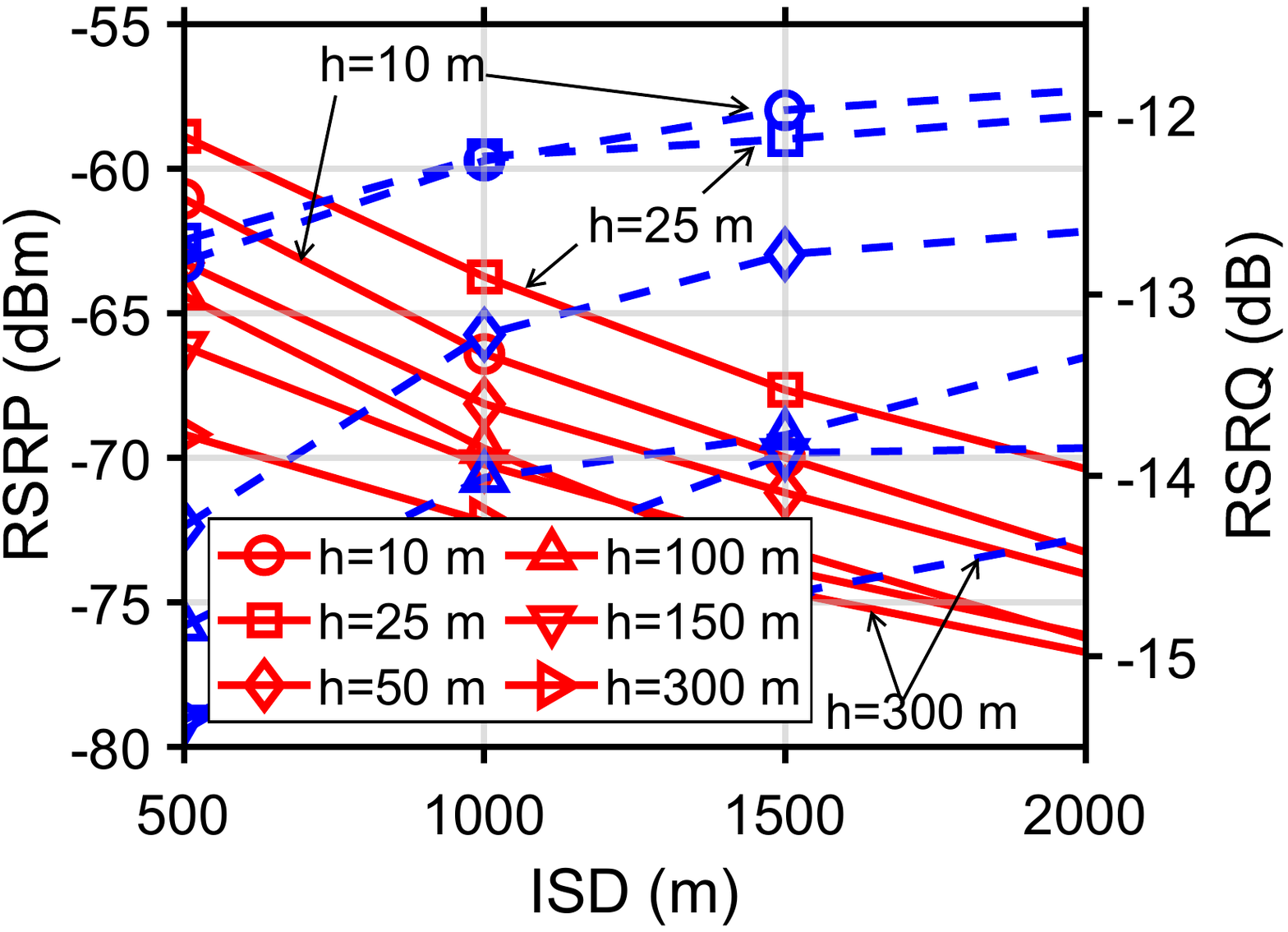}
}
\hfill
\subfloat[UMa scenario]{%
\includegraphics[width=0.48\linewidth]{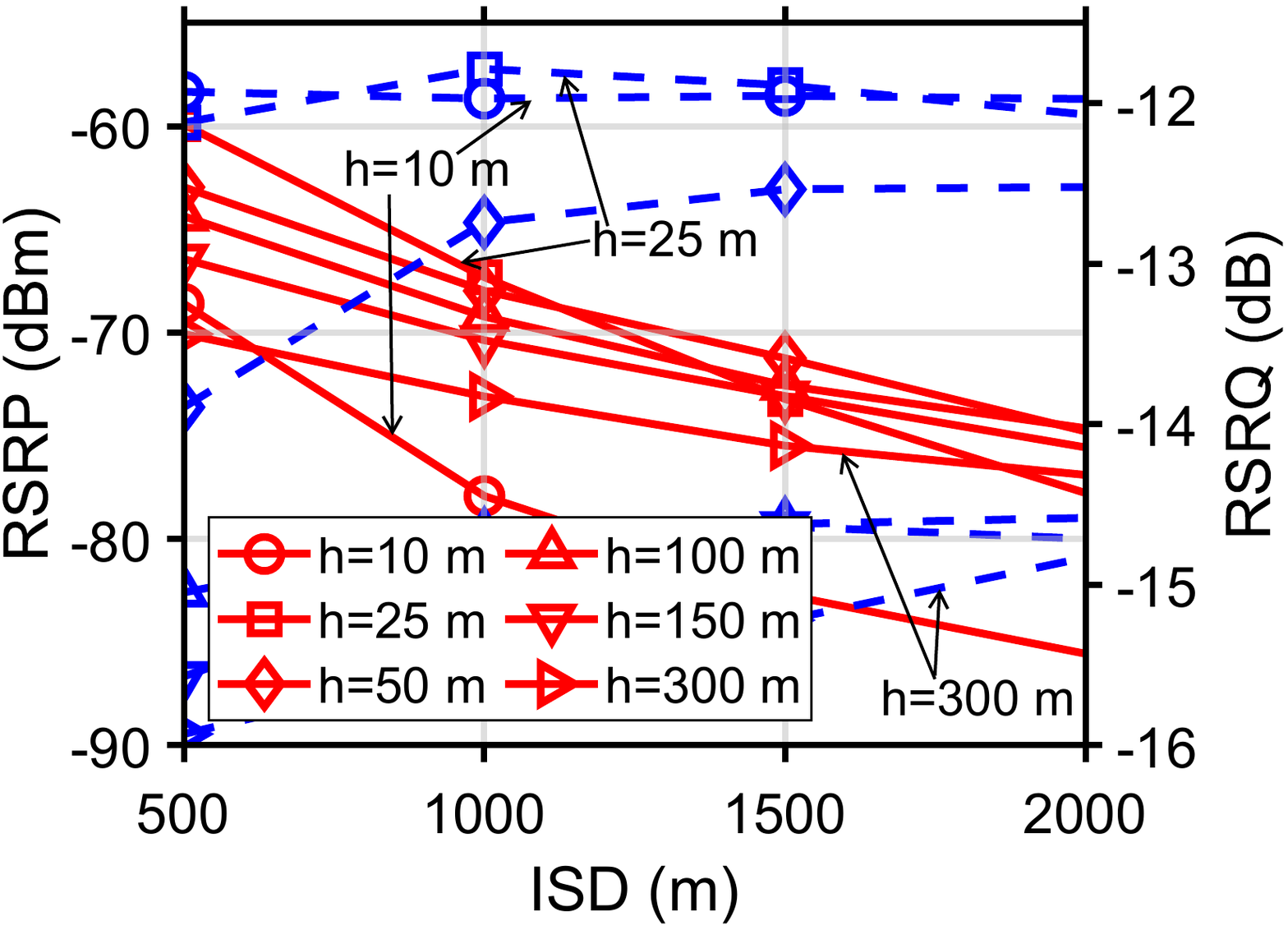}
}
\caption{RSRP (solid red) and RSRQ (dashed blue) versus ISD for multiple UAV altitudes.}
\label{fig:rsrp_rsrq_joint}
\vspace{-.2in}
\end{figure}

To understand the interplay between signal strength and interference, we jointly analyze RSRP and RSRQ as functions of ISD. In down-tilted cellular networks, UAVs are predominantly served through sidelobes rather than main lobes, which leads to moderate desired signal power while exposing the UAV to interference from multiple neighboring BSs. Fig.~\ref{fig:rsrp_rsrq_joint} shows that RSRP decreases with increasing ISD due to larger propagation distances. In contrast, RSRQ improves as ISD increases, since ICI reduces with greater BS separation. This opposing behavior highlights a key characteristic of sidelobe-driven aerial links, where interference, rather than signal strength, becomes the dominant factor. At higher altitudes, increased LOS visibility to multiple BSs further amplifies sidelobe interference, making the system interference-limited. These results demonstrate that coverage alone is insufficient to characterize UAV performance, and both signal strength and interference must be jointly considered.

\section{Conclusion}
\label{sec:conclusion}
In this paper, we present a 3GPP-compliant analysis of altitude-dependent performance for cellular-connected UAVs in 5G NR networks, incorporating realistic deployment geometry, antenna patterns, and probabilistic LOS/NLOS propagation. Our results show a clear transition from coverage-limited behavior at low altitudes to interference-limited operation at higher altitudes. While RSRP mainly reflects large-scale propagation and decreases with altitude and ISD, SINR degrades more rapidly due to increased exposure to inter-cell interference. In contrast, larger ISD improves SINR and RSRQ by reducing interference, despite weaker received power. RSRQ effectively captures this coverage–quality decoupling and highlights the growing impact of interference with altitude. These findings indicate that coverage-based metrics alone are insufficient for aerial users, and interference-aware design is critical for reliable UAV connectivity.
\bibliographystyle{IEEEtran}
\bibliography{references}

\end{document}